# A Spectral-model-independent Diagnostic of Phase-resolved Spectral Convergence in Swift J0243.6+6124

Bo-Yan Chen[1,2], Peng-Ju Wang[3,4], Shu Zhang[1], Ling-Da Kong[5], Long Ji[6], Shuang-Nan Zhang[1,2], Hua Feng[1], Yu-Peng Chen[1], Ming-Yu Ge[1,2], Jing-Qiang Peng[1,2], Qing-Cang Shui[1,2], and Wen-Zhong Li[1,2]
[1] State Key Laboratory of Particle Astrophysics, Institute of High Energy Physics, Chinese Academy of Sciences, Beijing 100049, People's Republic of China; bychen@ihep.ac.cn, szhang@ihep.ac.cn
[2] University of Chinese Academy of Sciences, Chinese Academy of Sciences, Beijing 100049, People's Republic of China
[3] Research Center of Astronomy, QingHai University, Xining, 810016, People's Republic of China; wangpj@qhu.edu.cn
[4] Department of Physics and Astronomy, QingHai University, Xining, 810016, People's Republic of China
[5] School of Physics, Nankai University, Tianjin 300071, People's Republic of China
[6] School of Physics and Astronomy, Sun Yat-Sen University, Zhuhai, 519082, People's Republic of China
*Received 2026 February 25; revised 2026 July 2; accepted 2026 July 19; published 2026 September 4*

## Abstract

We introduce a spectral-model-independent spectral-shape metric, $\delta_{\rm rms}$, to probe phase-resolved spectral evolution in the Galactic ultraluminous X-ray pulsar Swift J0243.6+6124. The $\delta_{\rm rms}$–luminosity relation reveals a reproducible convergence–redivergence behavior: spectra from two geometrically opposite pulse phases converge to near identity around $L_{\rm t}^* \sim 8.8 \times 10^{37}$ erg s$^{-1}$, with a mean shape difference of only ∼2%, and rediverge at higher luminosities, most prominently at higher energies. This strong phase-resolved spectral convergence coincides with a pulse-profile reversal and traces the same transition luminosities previously identified from broadband continuum evolution, including $L_{\rm t} \sim 4.5 \times 10^{37}$ erg s$^{-1}$. These results suggest a possible luminosity-driven reweighting between a broadly distributed surface-reprocessed/reflected component and an increasingly important direct accretion-column contribution, while establishing $\delta_{\rm rms}$ as a robust, spectral-model-independent diagnostic of radiation-geometry restructuring in bright accreting X-ray pulsars.

## 1. Introduction

Accretion-powered X-ray pulsars (XRPs) host strongly magnetized neutron stars in which the accretion flow is funneled toward the magnetic poles and radiates in highly anisotropic beaming patterns (M. M. Basko & R. A. Sunyaev 1976; P. A. Becker & M. T. Wolff 2007; A. Mushtukov & S. Tsygankov 2023). Pulse profiles therefore provide a sensitive probe of accretion geometry, radiative transfer, and strong-gravity effects, and often exhibit pronounced evolution with luminosity and photon energy (P. Meszaros & W. Nagel 1985; U. Kraus et al. 2003).

Pulse-profile modeling is frequently framed in terms of pencil and fan beams (M. M. Basko & R. A. Sunyaev 1976; A. N. Parmar et al. 1989). However, in bright XRPs the observed morphology is shaped by additional effects—including viewing geometry, occultation, and relativistic light bending. Recent simulations further suggest that in an accretion-column geometry a substantial fraction of photons can be scattered downward toward the stellar surface (L. Zhang et al. 2025), so that surface illumination and the associated surface-reprocessed/reflected emission become important (J. Poutanen et al. 2013; K. A. Postnov et al. 2015). As a result, the pulse profile is no longer determined by the intrinsic beaming pattern alone (H. Riffert & P. Meszaros 1988; A. A. Mushtukov et al. 2018). These effects introduce strong degeneracies in geometric interpretations, leaving the physical origin of pulse-profile evolution an open question.

In Be X-ray binary pulsars (BeXRPs), luminosity-dependent pulse-profile transitions are among the most striking manifestations of radiation-geometry evolution. Here, we define a specific subclass of such transitions as pulse-profile reversal (PPR), in which the dominant pulse peak is replaced by another peak as luminosity increases, typically following a single–double–single dominant pulse peak evolution. Peak exchange consistent with PPR was first clearly documented in EXO 2030+375 (A. N. Parmar et al. 1989), and similar peak exchange and/or single-/double-peaked transitions have been reported in other BeXRPs, including RX J0440.9+4431, 1A 0535+262, and Swift J0243.6+6124 (P. J. Wang et al. 2022; P. P. Li et al. 2023; M. M. Serim et al. 2023; P. Thalhammer et al. 2024; Y.-J. Du et al. 2025). Such behavior is often attributed to pencil–fan switching and/or interpreted as an empirical signature of entering the supercritical accretion regime.

At high luminosities, the accretion flow may form an extended accretion column above the neutron-star surface (M. M. Basko & R. A. Sunyaev 1976). Emission from the column can irradiate the stellar surface and generate a surface-reprocessed/reflected component, while relativistic light bending increases the visibility of the far side of the neutron star (H. Riffert & P. Meszaros 1988; J. Poutanen et al. 2013). In this picture, the PPR and the associated spectral evolution may reflect a luminosity-dependent redistribution between direct column emission and surface-reprocessed/reflected radiation, rather than a simple pencil–fan switching (K. A. Postnov et al. 2015; A. A. Mushtukov et al. 2018).

Phase-resolved spectroscopy provides a key approach to test these ideas, yet its interpretation often depends on the adopted continuum model, absorption treatment, and parameter degeneracies. We therefore introduce a new spectral-model-independent spectral-shape metric, $\delta_{\rm rms}$, which quantifies the root-mean-square fractional difference between two phase-resolved spectra relative to their mean spectrum; smaller values indicate more similar spectral shapes and enable direct tracking of luminosity-dependent spectral-shape evolution.

Swift J0243.6+6124, the only confirmed Galactic ultraluminous X-ray pulsar (ULXP), exhibits repeated outbursts spanning $\sim 10^{37}$–$10^{38}$ erg s$^{-1}$ (B.-Y. Chen et al. 2026) and shows a PPR, making it an ideal laboratory for investigating luminosity-driven geometric evolution in bright accreting XRPs (S. S. Tsygankov et al. 2018). Applying $\delta_{\rm rms}$ to its phase-resolved spectra, we uncover a pronounced nonmonotonic convergence–redivergence behavior with luminosity, in which $\delta_{\rm rms}$ reaches a deep minimum shortly after the PPR (down to $\sim$2%), revealing strong phase-resolved spectral convergence (PRSC). By examining the energy dependence of this behavior, particularly in higher-energy bands that are expected to be more sensitive to changes in the direct accretion-column contribution, we demonstrate that the PPR in Swift J0243.6+6124 is accompanied by a systematic restructuring of phase-resolved spectral shapes, and establish $\delta_{\rm rms}$ as a new, spectral-model-independent diagnostic of luminosity-driven geometric evolution in bright accreting XRPs.

## 2. Observations and Data Reduction

We use pointed Insight–HXMT observations of Swift J0243.6+6124 from three outbursts in 2018 (MJD 58125–58170), 2019 (MJD 58493–58523), and 2023 (MJD 60152–60181), spanning $L_{\rm bol} = 7.9 \times 10^{36}$–$1.2 \times 10^{38}$ erg s$^{-1}$ (2–150 keV; $d = 5.2$ kpc; C. A. L. Bailer-Jones et al. 2021; luminosities from B.-Y. Chen et al. 2026), fully covering the transition luminosities $L_{\rm t}$ and $L_{\rm t}^*$ (L. D. Kong et al. 2020; C. A. L. Bailer-Jones et al. 2021; B.-Y. Chen et al. 2026).

Photon arrival times are corrected to the solar system barycenter and for binary motion using HXMTDAS v2.06 and `hxbary`, adopting the orbital ephemeris from the Fermi Gamma-ray Burst Monitor (GBM).[7]

Pulse profiles are generated with `Stingray` (D. Huppenkothen et al. 2019) and aligned via Fourier-domain template matching with `fftfit` (J. H. Taylor 1992), and extracted in 1–10, 10–30, and 30–100 keV.

For spectroscopy, background-subtracted 2–100 keV spectra are grouped to a minimum of 100 counts per channel for the Low Energy X-ray Telescope (LE) and Medium Energy X-ray Telescope (ME), and 200 counts per channel for the High Energy X-ray Telescope (HE) (X. Cao et al. 2020; Y. Chen et al. 2020). Broadband (2–100 keV) spectra are used for phenomenological spectral fitting, whereas the $\delta_{\rm rms}$ metric is computed over 2–50 keV to avoid introducing large statistical uncertainties from energies above 50 keV in low-luminosity observations (X. Li et al. 2020; J.-Y. Liao et al. 2020). For phase-resolved fitting, spectra from $\phi = 0.0$–$0.5$ and $\phi = 0.5$–$1.0$ are modeled with `constant × TBabs × (bbodyrad + cutoffpl)`, allowing all parameters to vary independently between phases; the multiplicative constant accounts for cross-calibration differences.

[7] https://gammaray.nsstc.nasa.gov/gbm/science/pulsars/lightcurves/swiftj0243.html

## 3. Results

### 3.1. Pulse Profile Reversal

We examine the luminosity-dependent pulse evolution of Swift J0243.6+6124 using Insight–HXMT observations from all three outbursts, plotted together in luminosity order. The 10–30 keV band is shown in Figure 1, while the 1–10 keV and 30–100 keV results are presented in the Appendix.

Each pulse cycle is folded into 32 phase bins over $\phi \in [0, 1)$, with phase zero defined at the pulse minimum. The luminosity-ordered, row-wise normalized profiles reveal a gradual redistribution of emission between two broad phase intervals, $\phi = 0.0$–$0.5$ and $\phi = 0.5$–$1.0$.

Below $L_{\rm t}$, the profile is dominated by a broad peak near $\phi \sim 0.75$. Between $L_{\rm t}$ and $L_{\rm t}^*$, a peak near $\phi \sim 0.25$ strengthens and progressively overtakes the $\phi \sim 0.75$ peak, producing a pronounced double-peaked morphology and a clear exchange in dominance. Above $L_{\rm t}^*$, the reversal is complete, and the profile returns to a single-peak morphology stably dominated by the $\phi \sim 0.25$ peak. We refer to this systematic exchange in relative peak dominance as PPR.

To quantify this evolution, we compute the peak-to-peak and rms pulse fractions (P. J. Wang et al. 2020), as well as the integrated count rates and fractional contributions from the two opposite phase windows. The two previously identified transition luminosities, $L_{\rm t} \sim 4.5 \times 10^{37}$ erg s$^{-1}$ and $L_{\rm t}^* \sim 8.8 \times 10^{37}$ erg s$^{-1}$ (L. D. Kong et al. 2020; B.-Y. Chen et al. 2026), delineate three stages in these diagnostics. Below $L_{\rm t}$, both pulse fractions increase with luminosity, and the phase contributions are strongly asymmetric. Near $L_{\rm t}$, the pulse fractions exhibit clear breaks, and the two phase contributions become comparable. Between $L_{\rm t}$ and $L_{\rm t}^*$, the fractional contribution from the $\phi = 0.0$–$0.5$ window progressively exceeds that from $\phi = 0.5$–$1.0$, tracking the dominance exchange. Above $L_{\rm t}^*$, the $\phi = 0.0$–$0.5$ window consistently dominates, consistent with the stabilized single-peak morphology.

The 1–10 keV band shows similar behavior. In the 30–100 keV band, the lowest-luminosity profiles already appear double-peaked, and the dominance exchange is less sharply defined, although the post-$L_{\rm t}^*$ profiles are likewise dominated by the $\phi \sim 0.25$ peak.

### 3.2. Phase-resolved Spectral Convergence

To quantify the spectral evolution associated with the PPR, we introduce a spectral-model-independent metric that measures the luminosity-dependent similarity between phase-resolved spectra. For each observation, we extract two spectra from two geometrically opposite phase intervals, $\phi = 0.0$–$0.5$ and $\phi = 0.5$–$1.0$, and restrict the analysis to 2–50 keV to avoid the large statistical uncertainties above 50 keV at low luminosities (X. Li et al. 2020; J.-Y. Liao et al. 2020; Y.-J. Du et al. 2025).

The two spectra are rebinned onto a common linear energy grid using a flux-conserving scheme. To match the detector-dependent native resolution while maintaining statistical robustness, we adopt a piecewise grid within 2–50 keV consisting of 60 bins for LE, 30 bins for ME, and 10 bins for HE (X. Cao et al. 2020; Y. Chen et al. 2020; C. Liu et al. 2020).

Each rebinned spectrum is then converted into an area-fraction representation within the selected band. Specifically, for phase interval $k = 1, 2$ and rebinned energy bin $i$, the flux density is first converted to bin-integrated flux $F_{k,i} = y_{k,i}\Delta E_i$, where $\Delta E_i$ is the bin width. The band-integrated flux is $A_k = \sum_i F_{k,i}$, and the normalized area fraction is defined as

$$f_{k,i} = \frac{F_{k,i}}{A_k}, \tag{1}$$

so that $\sum_i f_{k,i} = 1$ over 2–50 keV. This normalization removes overall flux offsets and isolates spectral-shape information.

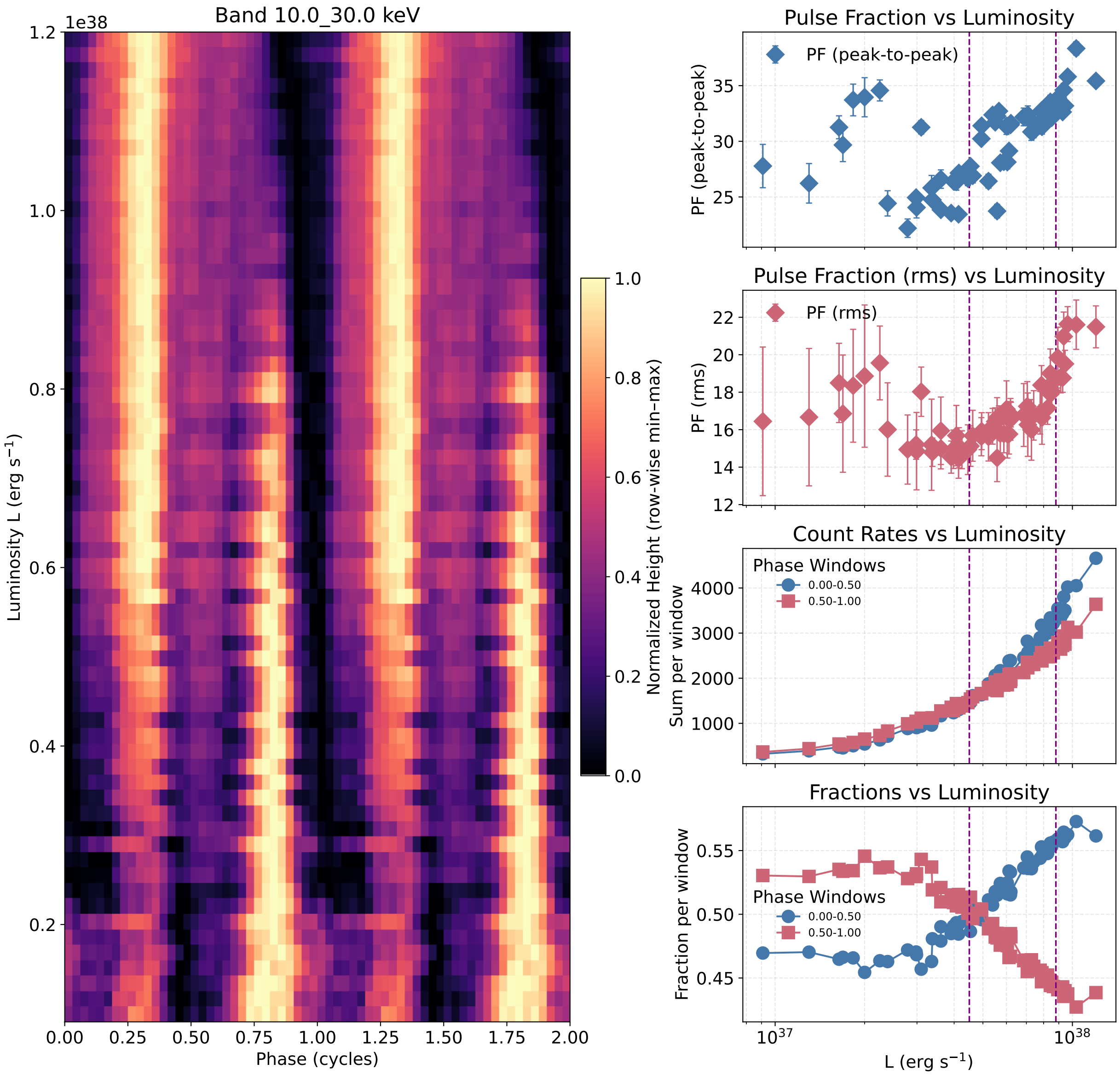

**Figure 1.** A comprehensive view of the pulse-profile evolution in the 10–30 keV band. Left: phase–luminosity matrix of the pulse profiles, where the vertical axis is the luminosity $L$ and the horizontal axis is the pulse phase. The color scale indicates the relative intensity normalized row-by-row using a min.–max. scaling (see the color bar). Right: four panels showing the luminosity dependence of pulse properties and the contributions of two geometrically opposite phase windows. From top to bottom: (1) the pulse fraction, $PF$; (2) the rms pulse fraction, $PF_{\rm rms}$; (3) the total count rate extracted from two opposite phase intervals (0.0–0.5 and 0.5–1.0) as a function of luminosity; and (4) the corresponding count fractions contributed by the two phase windows. The vertical purple dashed lines mark the previously identified spectral-transition luminosity $L_{\rm t}$ ( B.-Y. Chen et al. 2026) and $L_{\rm t}^*$ (L. D. Kong et al. 2020).

We define the $\Delta E$-weighted rms fractional deviation between the two normalized spectra as

$$\delta_{\rm rms} = \left[\frac{\sum_i \Delta E_i \left(\frac{f_{1,i} - f_{2,i}}{\bar{f}_i}\right)^2}{\sum_i \Delta E_i}\right]^{1/2}, \tag{2}$$

where $\bar{f}_i = (f_{1,i} + f_{2,i})/2$. By construction, $\delta_{\rm rms}$ is insensitive to overall normalization and provides a robust measure of spectral-shape differences. Although $\delta_{\rm rms}$ is independent of the adopted continuum spectral model, it still depends on the selected energy band, energy-binning scheme, and phase intervals.

Uncertainties are estimated via Monte Carlo resampling of the rebinned spectra. In each realization, flux densities are randomly drawn according to their statistical uncertainties, converted to area fractions, and used to recompute $\delta_{\rm rms}$. We adopt the median as the best estimate and the central 90% interval as the uncertainty.

Figure 2 shows a pronounced and reproducible nonmonotonic evolution of $\delta_{\rm rms}$ across the three outbursts. The two transition luminosities, $L_{\rm t}$ and $L_{\rm t}^*$ (L. D. Kong et al. 2020;

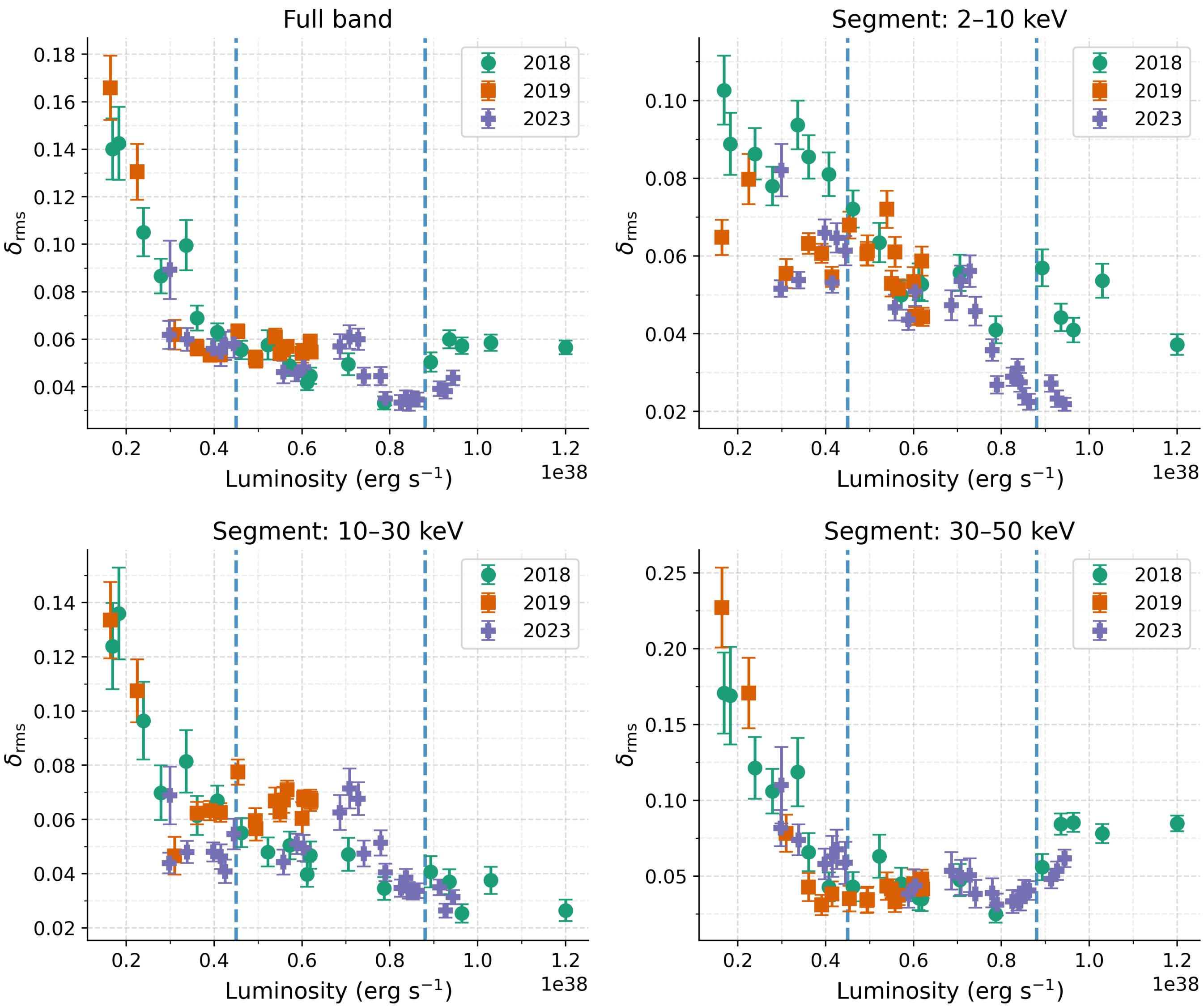

**Figure 2.** Evolution of $\delta_{\rm rms}$ with luminosity, computed between two geometrically opposite phase intervals ($\phi = 0.0$–$0.5$ and $\phi = 0.5$–$1.0$). The four panels show results for the full 2–50 keV band and three sub-bands (2–10, 10–30, and 30–50 keV). Error bars show 90% confidence intervals from Monte Carlo propagation. The vertical dashed lines mark the previously identified spectral-transition luminosity $L_{\rm t}$ and $L_{\rm t}^*$ (L. D. Kong et al. 2020; B.-Y. Chen et al. 2026).

B.-Y. Chen et al. 2026), naturally divide the spectral evolution into three regimes.

Below $L_{\rm t}$, $\delta_{\rm rms}$ decreases rapidly with increasing luminosity, indicating progressive convergence of the phase-resolved spectral shapes. A noticeable change in slope occurs near $L_{\rm t}$. Between $L_{\rm t}$ and $L_{\rm t}^*$, the convergence continues and reaches a deep minimum near $L_{\rm t}^*$, where $\delta_{\rm rms} \sim 2\%$, implying nearly identical spectral shapes between the two phase intervals. Above $L_{\rm t}^*$, $\delta_{\rm rms}$ increases again, marking a redivergence of the phase-resolved spectra.

We further divide the pulse cycle into 10 uniform phase bins and compute $\delta_{\rm rms}(\phi)$ relative to the reference peak phase $\phi = 0.8$–$0.9$ (dominant below the PPR), enabling a consistent peak-to-peak comparison across luminosities. Figure 3 shows representative $\delta_{\rm rms}(\phi)$ distributions at two luminosity levels on opposite sides of $L_{\rm t}^*$. At high luminosities, phase-to-reference shape differences are globally reduced, consistent with enhanced PRSC. Notably, the two PPR-related peak intervals ($\phi \sim 0.3$–$0.4$ and $\phi \sim 0.8$–$0.9$) exhibit a systematic inversion in relative similarity across $L_{\rm t}^*$: they correspond to the most similar phases below $L_{\rm t}^*$ but become the most discrepant phases above it. This inversion is reproducible across outbursts and reflects an intrinsic luminosity-driven restructuring of the phase-dependent spectral shapes.

To test whether the observed $\delta_{\rm rms}$ behavior depends on the adopted broad-phase windows, we performed an additional narrow-phase-bin check using the 2018 outburst, which provides the most complete luminosity coverage in our dataset. Specifically, we computed $\delta_{\rm rms}$ over 2–50 keV between the two 0.1-cycle phase bins associated with the PPR-related peaks, $\phi = 0.3$–$0.4$ and $\phi = 0.8$–$0.9$. As shown in Appendix Figure A1, this narrow-phase-pair measurement recovers the same luminosity-dependent convergence–redivergence pattern as the broad-window result, with the lowest $\delta_{\rm rms}$ values in the intermediate luminosity range and a renewed increase around and above $L_{\rm t}^*$. This test indicates that the observed PRSC is not an artifact of the particular broad-phase-window choice and is unlikely to be driven solely by pulse-profile evolution or by dilution within broad-phase windows.

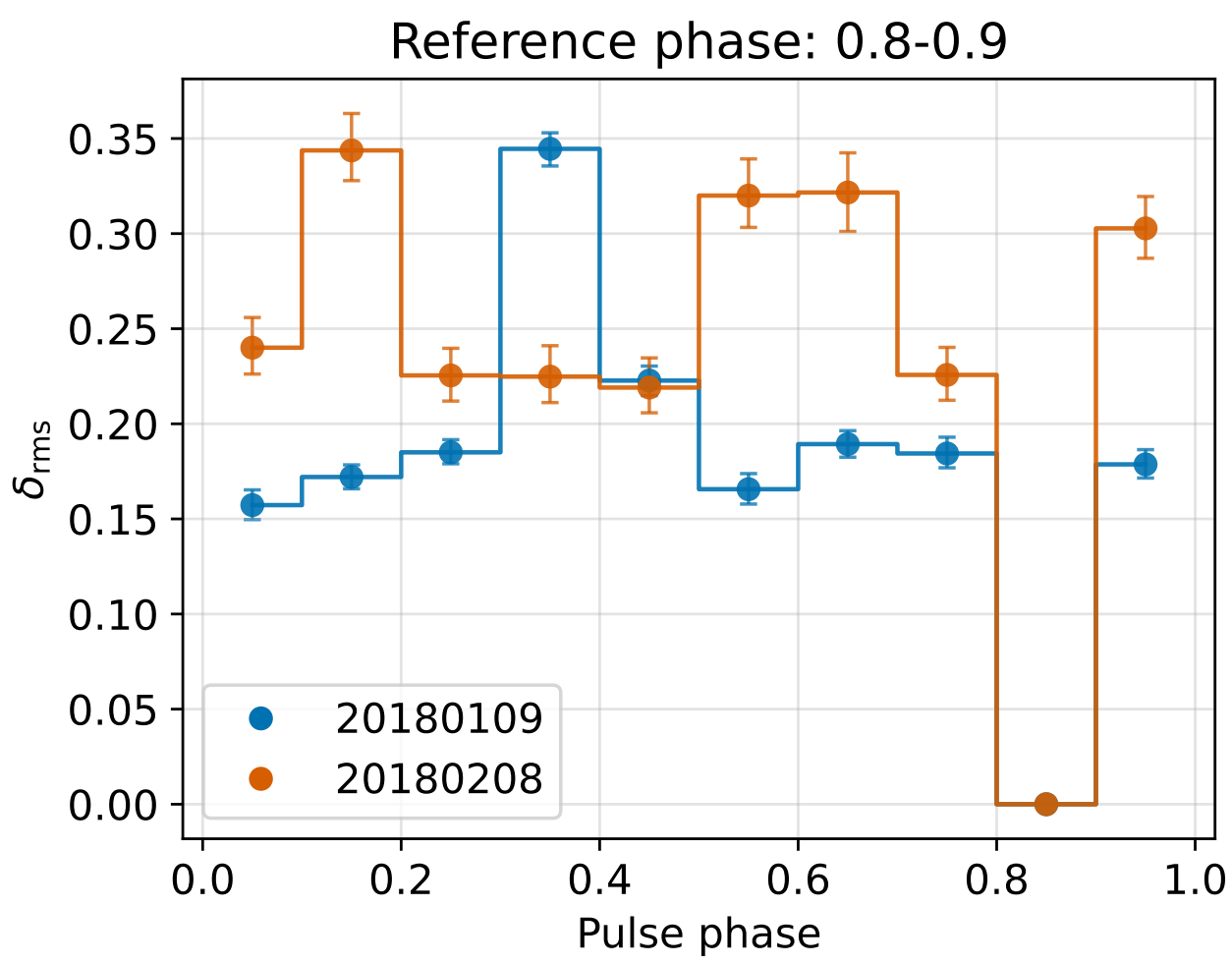

**Figure 3.** Representative $\delta_{rms}$ distributions as a function of pulse phase at two luminosity levels around $L_t^*$ (20180109: blue, $L \sim 1.03 \times 10^{38}$erg s$^{-1}$; 20180208: orange, $L \sim 3.36 \times 10^{37}$erg s$^{-1}$). The error bars indicate the measurements at the 90% confidence level.

We also tested whether phase-alignment uncertainties could affect this result. Small residual phase offsets mainly change photon assignment near phase-window boundaries and should add point-to-point scatter rather than create a coherent convergence–redivergence trend. Because $\delta_{rms}$ is computed from normalized spectral shapes, it is insensitive to pure pulse-height changes, and a half-cycle ambiguity would simply exchange the two compared spectra without changing $\delta_{rms}$. The consistent narrow-phase-pair resulting in Appendix Figure A1 further indicates that the observed PRSC is not driven by phase alignment.

### 3.3. Spectral Parameter Ratios

To test whether the $\delta_{rms}$ convergence reflects genuine spectral restructuring, we independently examine the luminosity evolution of phase-resolved spectral parameters. For each observation, spectra from $\phi = 0.0$–$0.5$ and $\phi = 0.5$–$1.0$ are fitted with `constant×TBabs×(bbodyrad + cutoffpl)` (B.-Y. Chen et al. 2026), and the ratios of best-fit parameters ($\phi = 0.0$–$0.5$ divided by $\phi = 0.5$–$1.0$) are computed.

Appendix Figure A2 shows the resulting parameter ratios as a function of luminosity. The normalization ratios of both the blackbody and cutoff power-law components systematically exceed unity above $L_t$, consistent with the dominance exchange identified in the pulse-profile analysis (Section 3.1). In contrast, the spectral-shape parameters display a different trend. The ratios of the blackbody temperature, photon index, and cutoff energy progressively approach unity between $L_t$ and $L_t^*$ and remain consistent with unity near $L_t^*$ within uncertainties. This multiparameter convergence coincides with the gradual decrease of $\delta_{rms}$ (Section 3.2), indicating that the two phase intervals are described by nearly identical continuum shapes at this luminosity. At higher luminosities, mild deviations from unity reappear in several parameters, consistent with the post-$L_t^*$ redivergence of $\delta_{rms}$.

The agreement between the $\delta_{rms}$ minimum and the model-dependent convergence of spectral-shape parameters supports the interpretation that the PRSC traces genuine luminosity-dependent restructuring of the projected radiation geometry, rather than a statistical artifact.

## 4. Discussion

J. Poutanen et al. (2024) modeled Swift J0243.6+6124 using IXPE polarization data and inferred a spin inclination of $i_p = 15°$–$40°$ and a magnetic obliquity of $\theta_p = 60°$–$80°$. Such a geometry allows the observer's line of sight to sweep across a broad range of magnetic viewing angles within one spin cycle, thereby producing strong phase-dependent spectral variations (P. Meszaros & W. Nagel 1985; H. Riffert & P. Meszaros 1988; P. A. Becker & M. T. Wolff 2007). As the accretion rate changes, both the phase-dependent spectral properties and the pulse profile can evolve, as shown in other strongly magnetized XRPs (A. A. Lutovinov et al. 2015). A. A. Mushtukov et al. (2018) theoretically constructed beam patterns determined jointly by direct accretion-column emission and radiation reflected from the neutron-star surface, and showed that changes in their relative weights with column height or luminosity can naturally reshape the pulse profile and provide a physical pathway for PPR. We therefore interpret the observed PRSC and PPR as luminosity-dependent reweighting between direct accretion-column emission and a broader phase-distributed surface-reprocessed/reflected component; this possible picture is schematically illustrated in Figure 4 as a three-stage geometric evolution.

For $L < L_t$, Swift J0243.6+6124 is generally considered to be in a subcritical accretion state (C. A. Wilson-Hodge et al. 2018; L. D. Kong et al. 2020). In this stage, the pulse profile is dominated by the peak near $\phi \approx 0.75$ and appears nearly single-peaked at low energies, consistent with pencil-beam-dominated emission expected in subcritical accretion. L.-D. Kong et al. (2022) reported a high-energy cyclotron absorption feature at $\sim$120–146 keV, which, if interpreted as an electron CRSF, corresponds to a strong magnetic field of $\sim 1.6 \times 10^{13}$ G. If this field strength were used directly to estimate the critical luminosity, a stable accretion column would be difficult to form within the luminosity range probed in this work. B.-Y. Chen et al. (2026) introduced a quadrupole component to construct a nondipolar magnetic-field configuration, in which the two magnetic poles have different field strengths, and proposed that $L_t$ may correspond to the luminosity at which a stable column starts to form at the weaker-field pole. In this picture, the large spectral-shape differences between opposite phases at low luminosities may reflect intrinsic differences in the local magnetic-field structure, accretion geometry, or angular emission pattern sampled at different phases.

In the intermediate luminosity range, $L_t < L < L_t^*$, the source undergoes a stable exchange of pulse-peak dominance: the peak near $\phi \approx 0.25$ overtakes the peak near $\phi \approx 0.75$, and the normalization ratios deviate from unity above $L_t$. At the same time, $\delta_{rms}$ continues to decrease, while the spectral-shape parameter ratios approach unity and reach their closest agreement near $L_t^*$. A conventional mixed pencil–fan beam picture can also produce similar pulse-profile evolution, in which the two reversing peaks are associated with intrinsically different beam components, namely pencil and fan beams (A. N. Parmar et al. 1989; P. Thalhammer et al. 2024). However, if this interpretation were sufficient, the two peaks would generally be expected to have distinct spectral shapes, rather than becoming nearly identical around $L_t^*$. A more natural explanation is that, after a stable column forms above $L_t$, a common reprocessed component distributed over a broad phase range grows and is progressively reweighted against the direct accretion-column emission. Downward scattering and relativistic light bending can redirect a substantial fraction of column photons toward the neutron-star surface, forming a broad-phase-distributed surface-reprocessed/reflected spectral pedestal that suppresses phase-dependent spectral-shape differences (J. Poutanen et al. 2013; K. A. Postnov et al. 2015; A. A. Mushtukov et al. 2018; I. D. Markozov & A. A. Mushtukov 2024; L. Zhang et al. 2025).

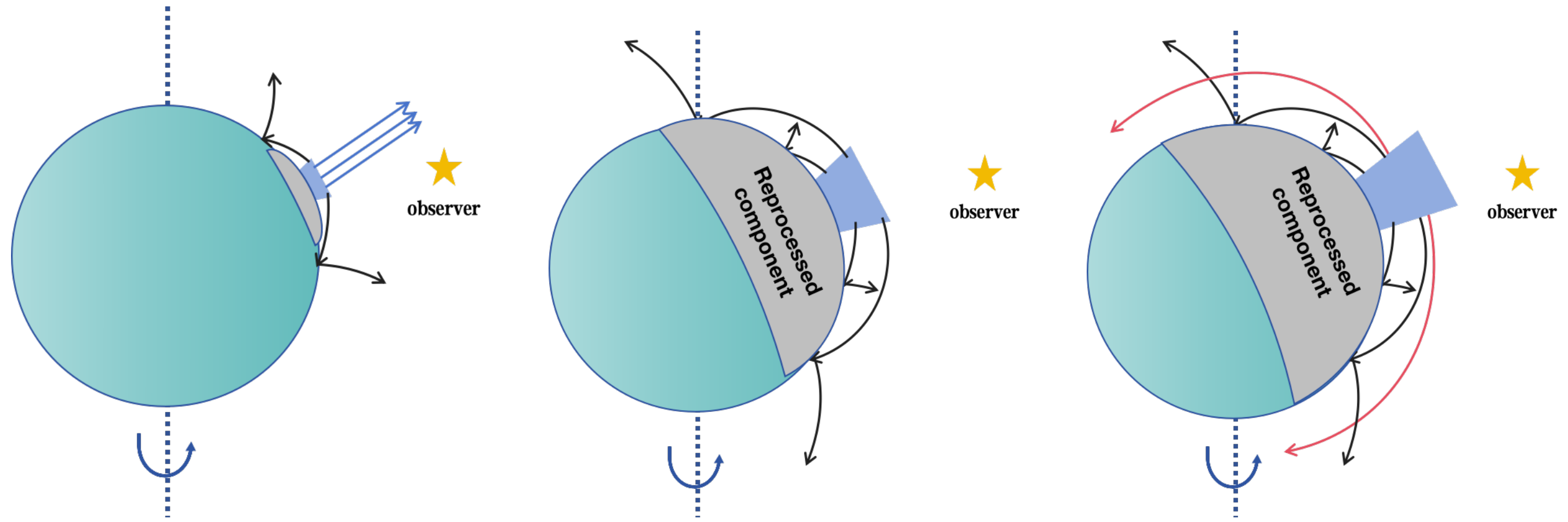

**Figure 4.** Schematic illustration of the luminosity-dependent geometric evolution in Swift J0243.6+6124. From left to right: (1) $L < L_t$: subcritical regime without a stable accretion column, where localized polar emission dominates, and strong phase asymmetry produces large spectral differences. (2) $L_t < L < L_t^*$: development of an extended accretion column and strong surface illumination, generating a broadly distributed reprocessed/reflected component that suppresses phase-dependent spectral differences and leads to convergence. (3) $L > L_t^*$: increasing visibility of direct column emission (including far-side contribution via light bending), breaking phase symmetry and driving spectral redivergence.

For $L > L_t^*$, $\delta_{rms}$ increases again, indicating renewed divergence of the phase-resolved spectral shapes. The 10-bin phase analysis shows that the two PPR-related peak phases reverse their relative similarity across $L_t^*$: they are among the most similar phases at low and intermediate luminosities, but become among the most discrepant phases at high luminosities. This symmetry breaking indicates that an additional direct accretion-column component is becoming stronger. This component may be associated with the farside column becoming more visible through light bending and becoming prominent near $\phi \approx 0.25$ (M. Sasaki et al. 2010; A. A. Mushtukov et al. 2018). In this framework, $L_t^*$ does not mark a new critical luminosity. Instead, it corresponds to the luminosity at which the surface-reprocessed component and the direct accretion-column emission are most nearly balanced, producing maximal spectral-shape convergence. Above this luminosity, the increasing direct column contribution drives the observed spectral redivergence.

The above picture may not be unique. A nondipolar two-pole accretion geometry, accretion-curtain scattering, and disk reflection/reprocessing may all participate in the PRSC and pulse-profile evolution to different degrees. Independent studies at higher luminosities support the presence of reprocessing structures with nonunique physical origins in Swift J0243.6+6124. NuSTAR phase-resolved reflection fitting found that the reflection-strength parameter in `relxilllp` varies with pulse phase, Insight-HXMT detected pulsed broad Fe-line emission, and NICER found soft X-ray line features in the super-Eddington state that may originate from disk or outflow reflection/reprocessing (S. D. Bykov et al. 2022; B. Chhotaray et al. 2024; Y. X. Xiao et al. 2024). However, these diagnostics mainly arise from X-ray features below 30 keV and therefore can hardly directly determine the physical origin of the PRSC observed here in the 30–50 keV band. We also consider the possible role of an optically thick accretion curtain or envelope. A. A. Mushtukov et al. (2017, 2019) showed that, at mass accretion rates comparable to those inferred for ultraluminous X-ray pulsars (ULXPs), the magnetospheric accretion flow can become closed and optically thick, modifying both the pulse profile and its phase dependence. This mechanism may be relevant to the peak of the 2017 giant outburst of Swift J0243.6+6124. However, the luminosity range analyzed in this work, $L_{bol} = 7.9 \times 10^{36}$–$1.2 \times 10^{38}$erg s$^{-1}$, is far below the regime in which a closed envelope is expected to dominate the observed emission. Therefore, disk/outflow reprocessing and an optically thick accretion curtain are unlikely to dominate the hard X-ray PRSC in the luminosity range studied here on their own.

For this hard band, the more natural reprocessing picture remains that hard column photons are reflected and angularly redistributed by the high-albedo atmosphere of the neutron-star surface. In a strongly magnetized neutron-star atmosphere, the albedo for hard photons can be high, and the reflected spectrum can be harder than the incident column radiation (J. Poutanen et al. 2013; K. A. Postnov et al. 2015; A. A. Mushtukov et al. 2018). Therefore, if the 30–50 keV spectral convergence is linked to reprocessing, surface reflection/reprocessing is more natural than disk, outflow, or accretion-curtain reflection/reprocessing. As the luminosity increases further, direct accretion-column Comptonized radiation regains a larger weight in the 30–50 keV band, naturally breaking the spectral-shape convergence near $L_t^*$ and producing the high-luminosity redivergence. Overall, $\delta_{rms}$ provides a sensitive and spectral-model-independent diagnostic of luminosity-driven restructuring of the projected radiation geometry in bright accreting XRPs.

## 5. Summary and Implications

We introduced a spectral-model-independent spectral-shape metric, $\delta_{rms}$, and found a reproducible convergence–redivergence pattern with luminosity in the Galactic ULXP

Swift J0243.6+6124. The phase-to-phase spectral-shape differences decrease with luminosity, reach a pronounced minimum of ~2% near $L_t^*$, and then increase again at higher luminosities. This PRSC is accompanied by the PPR and by consistent trends in the phase-resolved spectral-parameter ratios, revealing a three-stage luminosity-dependent evolution separated by $L_t \sim 4.5 \times 10^{37}$ and $L_t^* \sim 8.8 \times 10^{37}$ erg $s^{-1}$. We emphasize that $\delta_{rms}$ measures spectral-shape differences after projection along the observer's line of sight. Since both direct column emission and reprocessed/reflected emission can be intrinsically angle-dependent, spectral convergence does not necessarily imply identical intrinsic spectra at different phases, nor does it require a single emission component to dominate.

These results indicate that the luminosity scales inferred from continuum evolution, pulse morphology, and phase-resolved spectral-shape convergence are physically connected. The broadband spectral-hardness turnover in Swift J0243.6+6124 also occurs near $L_t^*$ (L. D. Kong et al. 2020). As suggested by K. A. Postnov et al. (2015), such a transition may be associated with a decreasing contribution from surface reflection and a growing importance of direct column emission. In this framework, the observed PRSC and its subsequent redivergence can be understood as a luminosity-dependent reweighting between direct accretion-column radiation and a broader phase-distributed surface-reprocessed/reflected component, although this interpretation is not unique.

Our results also have implications for empirical critical-luminosity diagnostics in bright accretion-powered XRPs. Broadband hardness breaks and luminosity-dependent pulse-profile changes remain useful indicators that the system has entered a regime where an extended accretion column, strong surface illumination, and geometric restructuring become important. However, such observables do not necessarily mark the exact onset of stable column formation, nor do they uniquely imply a discrete switch of intrinsic beaming patterns. By quantifying how phase-dependent spectral-shape differences are suppressed and reemerge with luminosity, $\delta_{rms}$ provides a complementary and relatively spectral-model-independent diagnostic of luminosity-driven restructuring of the projected radiation geometry in bright accreting XRPs.

## Acknowledgments

This work made use of data from the Insight-HXMT mission, a project funded by the China National Space Administration (CNSA) and the Chinese Academy of Sciences (CAS). This work is supported by China's Space Origins Exploration Program, the National Key R&D Program of China (2021YFA0718500), and the National Natural Science Foundation of China (NSFC) under grant Nos. 12025301 and 12333007. It is also partially supported by the International Partnership Program of the Chinese Academy of Sciences (grant No. 113111KYSB20190020). We acknowledge support from China's Space Origins Exploration Program (eXTP). Lingda Kong acknowledges support from the Fundamental Research Funds for the Central Universities (grant No. 010-63263126).

## Appendix
## Additional Pulse and Spectral Analysis Figures

This appendix presents supplementary figures that support the phase-resolved timing and spectral analyses discussed in the main text. Figure A1 shows the narrow-phase-bin robustness check used to test the dependence of $\delta_{rms}$ on the phase-window selection. Figure A2 presents the luminosity evolution of phase-resolved spectral-parameter ratios, and Figure A3 compares the pulse-profile evolution in the soft and hard energy bands.

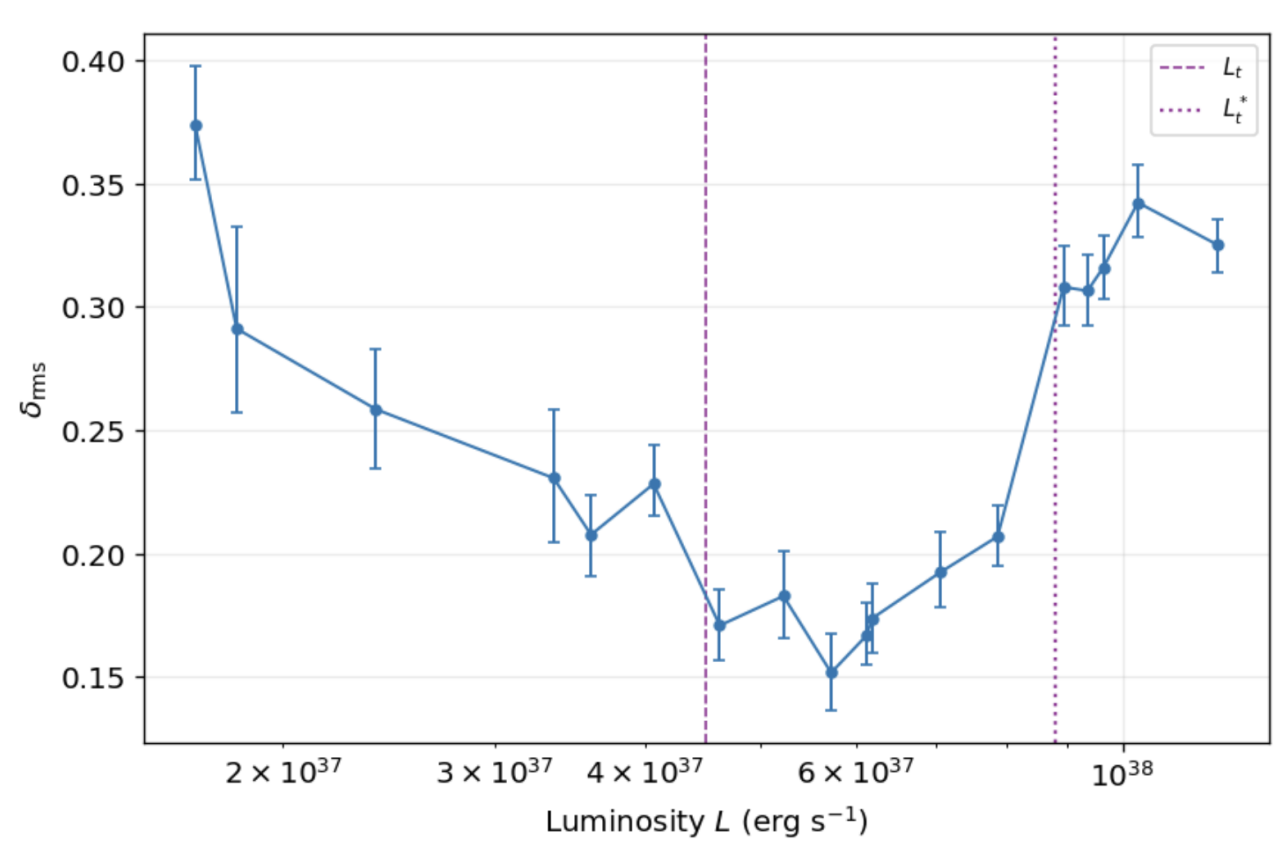

**Figure A1.** Narrow-phase-bin robustness check for the 2018 outburst, computed over 2–50 keV. The plotted $\delta_{rms}$ values are computed between the two 0.1-cycle phase bins associated with the PPR-related peaks, $\phi = 0.3$–0.4 and $\phi = 0.8$–0.9. The vertical dashed and dotted magenta lines mark $L_t$ and $L_t^*$, respectively. The narrow-phase-pair measurement shows a luminosity-dependent convergence–redivergence trend similar to that obtained from the broad-phase windows, supporting that the PRSC is not produced by the particular broad-window definition alone.

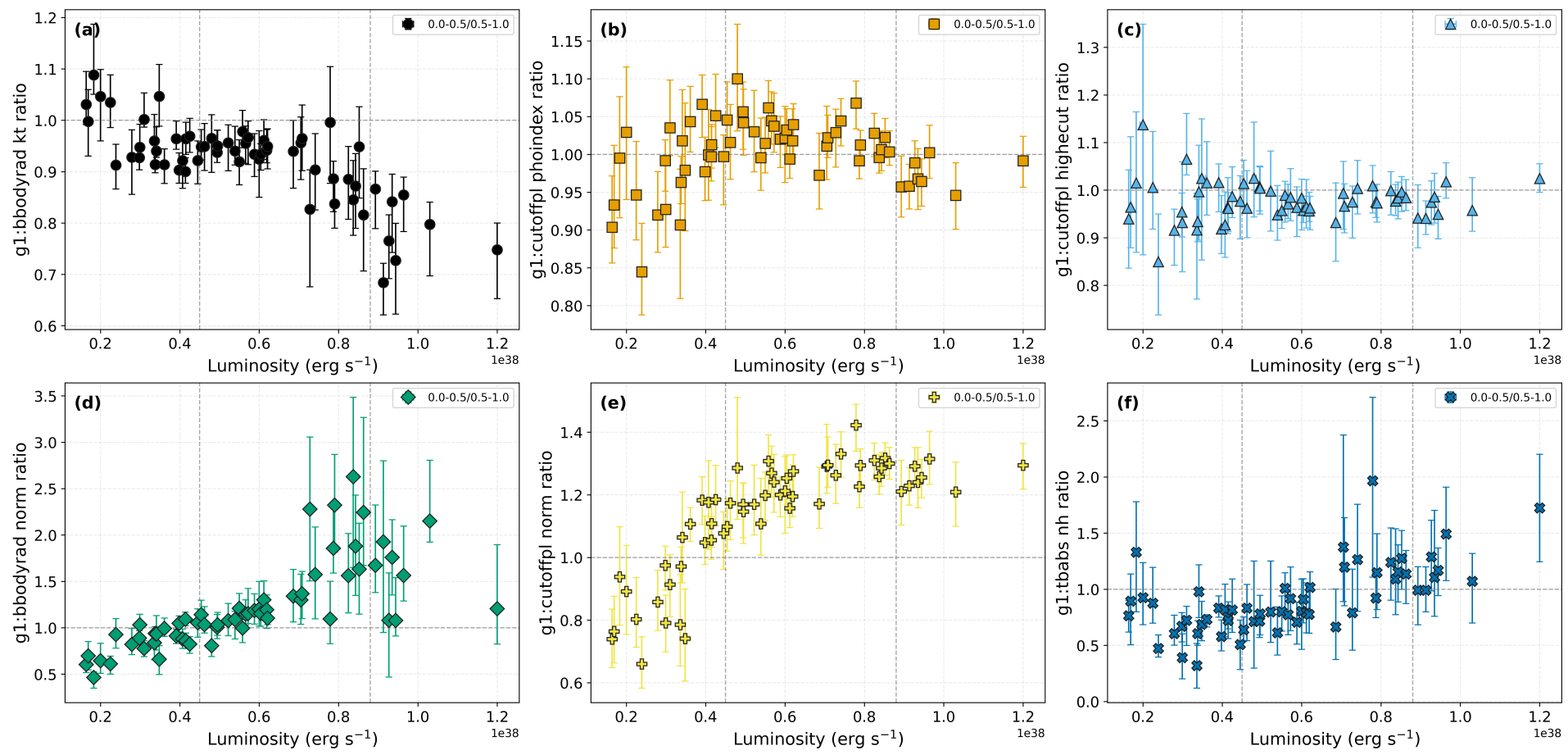

**Figure A2.** Luminosity evolution of phase-resolved spectral parameter ratios between $\phi$ = 0.0–0.5 and $\phi$ = 0.5–1.0, derived from fits with `constant×TBabs×(bbodyrad + cutoffpl)`. Panels (a)–(f) show the ratios of $kT$, blackbody normalization, $E_{\rm cut}$, cutoff power-law normalization, photon index Γ, and $N_{\rm H}$, respectively. The horizontal dashed line marks unity, and the vertical dashed lines indicate $L_{\rm t}$ and $L_{\rm t}^*$ (L. D. Kong et al. 2020; B.-Y. Chen et al. 2026).

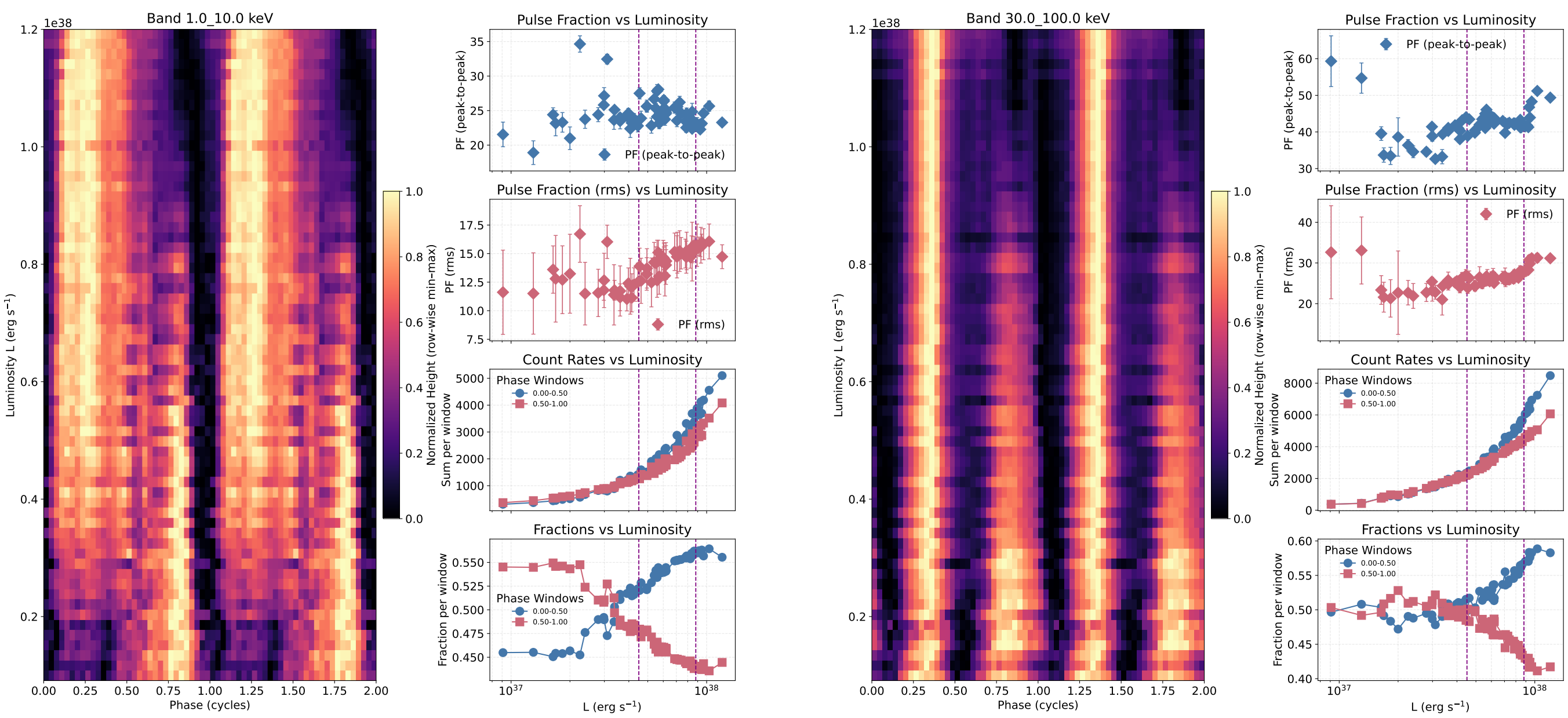

**Figure A3.** Same as Figure 2, but for the 1–10 keV and 30–100 keV bands.

## ORCID iDs

Bo-Yan Chen https://orcid.org/0009-0008-0518-6795
Peng-Ju Wang https://orcid.org/0000-0002-6454-9540
Shu Zhang https://orcid.org/0000-0003-3972-2564
Ling-Da Kong https://orcid.org/0000-0003-3188-9079
Long Ji https://orcid.org/0000-0001-9599-7285
Shuang-Nan Zhang https://orcid.org/0000-0001-5586-1017
Hua Feng https://orcid.org/0000-0001-7584-6236
Yu-Peng Chen https://orcid.org/0000-0001-8768-3294
Ming-Yu Ge https://orcid.org/0000-0002-3776-4536
Jing-Qiang Peng https://orcid.org/0000-0002-5554-1088
Qing-Cang Shui https://orcid.org/0000-0001-5160-3344
Wen-Zhong Li https://orcid.org/0009-0001-3113-586X